\documentclass[letterpaper]{jpconf}
\usepackage{graphicx}
\usepackage{bm}
\begin{document}
\title{Plans for a Neutron EDM Experiment at SNS}

\author{Takeyasu M. Ito\footnote{For the nEDM Collaboration}}

\address{Los Alamos National Laboratory, H846, PO Box 1663, Los
  Alamos, NM, 87545, USA}

\ead{ito@lanl.gov}

\begin{abstract}
The electric dipole moment of the neutron, leptons, and atoms provide a
unique window to Physics Beyond the Standard Model. We are currently
developing a new neutron EDM experiment (the nEDM
Experiment)~\cite{nEDM}. This experiment, which will be run at the
8.9~\AA\ Neutron Line at the Fundamental Neutron Physics Beamline
(FNPB) at the Spallation Neutron Source (SNS) at the Oak Ridge
National Laboratory, will search for the neutron EDM with a
sensitivity two orders of magnitude better than the present limit. In
this paper, the motivation for the experiment, the experimental
method, and the present status of the experiment are discussed.
\end{abstract}

\section{Introduction}
A nonzero permanent electric dipole moment (EDM) of a nondegenerate
state of a system with spin $J \neq 0$ violates the invariance under
time reversal as well as the invariance under parity operation. The
violation of time reversal invariance implies a violation of
invariance under $CP$ operation (combined operations of parity and
charge conjugation) through the $CPT$ theorem.

Within the standard model (SM), in the electroweak sector $CP$
symmetry is broken by the complex phase ($\delta_{\rm KM}$) in the CKM
quark mixing matrix (the KM mechanism). To date, in laboratory
measurements $CP$ violation has only been observed in $K$ and $B$
meson decays and the SM description of the $CP$ violation agrees with
all the laboratory measurements to date. However, the question remains
whether or not there are additional sources of $CP$ violation from new
physics. Indeed, almost all extensions of the SM imply that there are
such additional sources. Moreover, $CP$ violation is one of the
necessary conditions for the matter-antimatter asymmetry observed in
the Universe, and the SM and its description of $CP$ violation fail to
accommodate the observed asymmetry. This discrepancy suggests that
there are additional sources of $CP$ violation beyond that in the SM.

The current efforts to search for an electric dipole moment (EDM) of
the neutron are motivated by the following two observations, which
make the neutron EDM (as well as the EDM of other particles) an ideal
place to search for new sources of $CP$ violation:
\begin{enumerate}
\item In the KM mechanism, $CP$ violation only occurs in quark flavor
changing processes to the lowest order. Therefore, the EDM due to this
SM source of $CP$ violation is small; calculations predict it to be
$|d_n| \sim 10^{-32}-10^{-31}$~$e\cdot$cm, several orders of magnitude
smaller than the sensitivity possible with any experiment being
considered at present.
\item Most extensions of the SM naturally produce larger EDMs because
of additional $CP$ violating phases associated with additional
particles introduced in the model.
\end{enumerate}

Furthermore, the SM has another source of $CP$ violation. This
is a term in the QCD Lagrangian, the so-called $\theta$ term,
\begin{equation}
{\cal L}_{QCD} = {\cal L}_{QCD,\theta=0} 
+ \frac{\theta g_s^2}{32\pi^2}G_{\mu\nu}\tilde{G}^{\mu\nu},
\end{equation}
which explicitly violates $CP$ symmetry because of the appearance of
the product of the gluonic field operator $G$ and its dual $\tilde
G$. Since $G$ couples to quarks but does not induce flavor change,
$d_n$ is much more sensitive to $\theta$ than it is to $\delta_{\rm
KM}$. Thus measurement of $d_n$ determines an important parameter of
the SM. Calculations have shown that $d_n \sim {\cal O}
(10^{-16}\theta)$~$e\cdot$cm~\cite{POS99}. The observed limit on
$d_n$~\cite{BAK06} provide a limit $\theta < {\cal O}(10^{-10})$. A
similar limit can be obtained from the observed limit on the EDM of
the $^{199}$Hg atom~\cite{ROM01}. On the other hand, the natural scale
suggests rather that $\theta \sim {\cal O}(1)$. The puzzle of why the
value for $\theta$ is so small is called the strong $CP$
problem. Peccei and Quinn proposed a solution to this problem, in
which the strong $CP$ is a conserved quantity. This solution, however,
predicts the existence of a light pseudoscalar, called the axion. No
axions have yet been observed despite extensive searches.  It is
important to establish experimentally if $\theta$ has a very small
but finite value or $\theta$ is zero in order to shed light on this
problem.

The current upper limit on the neutron EDM $d_n$ comes from a
measurement performed at Institut Laue Langevin (ILL)~\cite{BAK06} and
is $|d_n| < 2.9 \times 10^{-26}$~$e$~cm (90\% C.L.). With many
theories predicting values lying within the six orders-of-magnitude
window between the current limit and the value allowed by the SM,
neutron EDM experiments that explore the next two orders of magnitude
would make a significant contribution to the search for new physics
that complements the information to be gained from the Large Hadron
Collider (LHC).

We are currently developing a new neutron EDM experiment (the nEDM
Experiment)~\cite{nEDM}. This experiment, which will be run at the
8.9~\AA\ Neutron Line at the Fundamental Neutron Physics Beamline
(FNPB) at the Spallation Neutron Source (SNS) at the Oak Ridge
National Laboratory, will search for the neutron EDM with a
sensitivity two orders of magnitude better than the present
limit. 
A non-zero EDM will be a
clear signal of physics beyond the SM, while a two-order-of-magnitude
improvement on the limit will provide a significant challenge to many
of the models of extensions of the SM.
  
\section{Experimental Method}
\subsection{Overview}
In the presence of a nonzero EDM, there is an interaction between the
EDM and a static electric field that is analogous to the interaction
between the magnetic moment and a static magnetic field. The
Hamiltonian is
\begin{equation}
\label{eq:hamiltonian}
H=-(\bm{\mu}_n \cdot \bm{B} + \bm{d}_n\cdot \bm{E}),
\end{equation}
where $\bm{B}$ and $\bm{E}$ are the applied static magnetic and
electric fields, $\bm{\mu}_n$ is the magnetic moment, and $\bm{d}_n$
is the EDM of the neutron.  The $\bm{d}_n\cdot \bm{E}$ interaction
causes the neutron spin to precess when a neutron is placed in a
static electric field, just like the $\bm{\mu}_n\cdot\bm{B}$
interaction does when the neutron is subject to a static magnetic
field.  Therefore, the EDM can be measured by looking for the change
in spin precession frequency of the neutron associated with a reversal
of $\bm{E}$ relative to $\bm{B}$. More specifically, the value of the
EDM is given by
\begin{equation}
d_n = \frac{h\Delta \nu}{4E},
\end{equation}
where $h$ is the Planck constant, $\Delta \nu$ is the change in the
precession frequency associated with a reversal of $\bm{E}$ relative
to $\bm{B}$, and $E$ is the strength of the applied static electric
field.

In a typical experimental arrangement, a sample of polarized neutrons
is introduced into a volume where uniform magnetic and electric fields
are applied, and the neutrons are let precess for a certain amount of
time.  In such experiments based on spin precession measurements, the
statistical uncertainty ($\delta d_n$) is ultimately given by the
uncertainty principle and is~\cite{KHR97}
\begin{equation}
\delta d_n = \frac{\hbar}{2ET\sqrt{mN}},
\end{equation}
where $T$ is
the time for which the neutron spin is let precess, and $m$ represents
the number of separate complete measurements of the $N$ neutrons. 

For experiments in which stored ultracold neutrons (UCNs)\footnote{
Ultracold neutrons are neutrons with total kinetic energy less than
the effective potential $U_F$ presented by a material boundary. These
neutrons, therefore, can be confined in a material bottle.  Typically
$U_F\sim 200$~neV, which corresponds to velocities of order 5~m/s,
wavelengths of order 500~\AA\, and an effective temperature of order
2~mK. See, for example Ref.~\cite{GOL91}} are used,\footnote{The use
of UCNs is important in suppressing the motional magnetic field
effects~\cite{SAN64}, which were a major limitation in early
experiments with a cold neutron beam.} where $N$ neutrons are stored
for a time $T_{\rm store}$, in a given total time $t$ the number of
measurements $m$ is inversely proportional to the storage time
$T_{\rm store}$ ($m\sim t/T$). Therefore,
\begin{equation}
\delta d_n \propto \frac{1}{E\sqrt{NT_{\rm store}}}.
\end{equation}
It is therefore obvious that in designing an EDM experiment, it is
important to maximize $E$, $N$,and $T_{\rm store}$.  In the most
recent ILL experiment~\cite{BAK06}, $E=10$~kV/cm, $T_{\rm store}=130$~s,
and $N=1.4\times 10^4$.

In the nEDM Experiment we intend to achieve a two orders of magnitude
in sensitivity by adopting the method proposed by Golub and
Lamoreaux~\cite{GOL94}, which is qualitatively different from the
methods adopted in previous experiments (for the history of neutron
EDM experiments, see for example Ref.~\cite{KHR97}). The overall
strategy can be summarized as follows:
\begin{itemize}
\item The experiment is performed in a bath of superfluid $^4$He. 
\item UCNs are produced locally in the measurement cells via the
  downscattering of 8.9~\AA\ cold neutrons in the superfluid helium
  (superthermal process).
\item A dilute admixture of polarized $^3$He atoms is introduced in
  the bath of superfluid $^4$He and is used as the co-magnetometer.
\item The polarized $^3$He atoms are also used as the neutron spin
  analyzer; by observing the spin-dependent $n+^3{\rm He}\to p+t$
  reaction, the difference between the neutron and $^3$He precession
  frequencies can be measured.
\item The $^3$He precession frequency is determined by directly
  measuring the change in the magnetic field caused by the rotating
  magnetization of the $^3$He atoms using SQUID magnetometers.
\end{itemize}
With this strategy, we expect to achieve $E=50$~kV/cm, $T_{\rm
store}\sim 500$~s, and $N\sim 10^6$, resulting in a
two-orders-of-magnitude improvement in sensitivity. Some of the
important features of this strategy will be described in more details
below.

\subsection{UCN Production}
\begin{figure}
\begin{center}
\includegraphics[width=6cm]{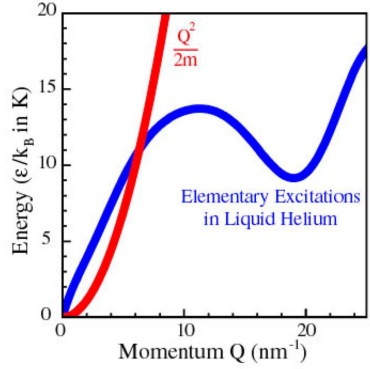}
\end{center}
\caption{The dispersions curve for the elementary excitation in
  superfluid $^4$He (blue) and the dispersion curve for the free
  neutron (red). The two curves cross at $Q=0$ and $Q=Q^*$, which
  corresponds to a neutron wavelength of 8.9~\AA\ or an energy of 12
  K.}
\label{fig:superthermal}
\end{figure}

In previous EDM experiments with stored UCNs, UCNs produced elsewhere
were transported to the experimental apparatus and were stored in the
measurement cells. The number of UCNs suffer from the loss in the
extraction of UCNs from the UCN source and during the transport from
the source to the experiment. These problems can be overcome if UCNs
can be generated directly in the EDM measurement cell. It is in fact
possible using the superthermal process~\cite{GOL77} in superfluid
liquid $^4$He with 8.9~\AA\ neutrons delivered to the measurement
cell. This mechanism is illustrated in Fig.~\ref{fig:superthermal}.
Shown in the figure are the dispersion curve of the elementary
excitation in superfluid $^4$He and that of the free neutron, which
cross at $Q=0$ and $Q=Q^*$, which corresponds to a neutron wavelength
of 8.9~\AA\ or an energy of 12~K. Therefore, neutrons with a
wavelength of 8.9~\AA\ can give all their energy and momentum to the
elementary excitations in superfluid helium and become ultracold
neutrons. This process is called ``downscattering.'' The reverse
process ``upscattering'' is highly suppressed, because there are hardly
any elementary excitations with energy of $E^*=12$~K ($Q=Q^*$) when the
temperature of the liquid helium is $T<0.5$~K (suppressed by
$\exp(-E^*/kT)$). The production of UCNs using this method has been
experimentally demonstrated~\cite{AGE78,GOL83}.

Since $^4$He has zero neutron absorption, UCN can be store in the bath
of superfluid $^4$He until $\beta$ decay, wall absorption, or
upscattering occurs. Our goal is to achieve a 500~s lifetime for the
stored UCNs. With this and the expected neutron flux at the 8.9~\AA\
Line at FNPB (see Sec~\ref{sec:fnpb}), we expect to have $N\sim 10^6$
per measurement. We plan to use a beam of polarized neutrons to
generate UCNs polarized. 

There are other advantages to performing an EDM experiment directly in
a bath of liquid helium. A dilute admixture of polarized $^3$He atoms
can be introduced into the bath, and can serve as a comagnetometer and
neutron spin analyzer as we will see below. In addition, because of
the excellent dielectric properties of liquid helium, it is expected
that a higher electric field can be applied in liquid helium than in
vacuum. 

\subsection{Magnetic Field Measurement with a $^3$He Comagnetometer}
An EDM of $10^{-28}$~$e\;$cm would produce a relative change of $\sim
5\times 10^{-9}$~Hz in precession frequency on reversal of $\bm{E}$
with respect to $\bm{B}$ for $|\bm{E}| = 50$~kV/cm. This frequency
shift corresponds to a change in magnetic field of about $2\times
10^{-12}$~gauss. Obviously extreme care needs to be taken to ensure
and monitor the temporal stability and spatial uniformity of the
magnetic field in order to minimize any possible systematic effects.

We will use polarized $^3$He atoms as comagnetometer as a means to
address this issue.. A comagnetometer is a polarized atomic species
within the same storage volume as the neutrons, which provides a
nearly exact spatial and temporal average of the magnetic field
affecting the neutrons over the storage period. The effectiveness and
importance of the use of comagnetometer have recently been demonstrated
experimentally in the ILL experiment~\cite{BAK06,HAR99}.

We will introduce a dilute admixture of polarized $^3$He atoms into
the superfluid $^4$He. The fractional density of $^3$He will be
$x=N(^3{\rm He})/N(^4{\rm He})\sim 10^{-10}$ (This is determined from
the neutron lifetime in the helium bath, as discussed below). $^3$He
is a diamagnetic atom and the EDM of $^3$He is negligible due to the
shielding from the two bound electrons. The precession of $^3$He atoms
will be determined by directly measuring the change in the magnetic
field caused by the rotating magnetization of the $^3$He atoms using
SQUID magnetometers.

\subsection{Measurement of the Neutron Precession Frequency}
The $^3$He atoms also serves as an analyzer of the neutron precession
frequency. The cross section for neutron absorption by $^3$He is
strongly spin dependent: 59~b for spins aligned and 11~kb for spins
opposite at thermal neutron energy (25.3~meV). The cross sections for
both scale inversely with neutron velocity. Thus if the $^3$He
concentration is adjusted to $10^{12}$~atoms/cc ($x=N(^3{\rm
He})/N(^4{\rm He})\sim 10^{-10}$), then the neutrons are only absorbed
when the neutron spin is opposite to the $^3$He spin. Therefore the
rate of neutron absorption on $^3$He is proportional to
\begin{equation}
1-P_3 P_n \cos[(\gamma_3 - \gamma_n)B_0 t],
\end{equation}
where $P_3$ and $P_n$ are the polarizations of the $^3$He and the
neutrons respectively, and $\gamma_3$ and $\gamma_n$ are the
gyromagnetic ratios of the $^3$He and the neutrons respectively. 

When a neutron is absorbed on a $^3$He atom, the reaction products are
a proton and a triton that share 764~keV of energy. When the reaction
products travel in liquid helium, they produce scintillation light in
the hard ultraviolet ($\sim 80$~nm). If the measurement cells are
coated with a wavelength shifter, the scintillation light is converted
to the blue which may be detected by photomultiplier tubes (PMT). 

By measuring the rate of scintillation light, the difference between
the neutron precession frequency and the $^3$He precession frequency
can be determined. Note that the $^3$He precession frequency is
measured using the SQUID magnetometer. The signature of a neutron EDM
would appear as shift in the neutron$-^3$He precession frequency
difference corresponding to the reversal of the $\bm{E}$ relative to
$\bm{B}$ with no corresponding change in $^3$He precession. 

\subsection{Dressed Spin Technique}
In addition to the use of SQUID magnetometer to monitor the polarized
$^3$He, the so-called ``dressed-spin technique'' can be used to look
for the signature of the neutron EDM. This is based on the fact that
in the presense of a strong oscillating magnetic field, the magnetic
moment of a particle is modified or ``dressed'', yielding an effective
gyromagnetic ratio given by
\begin{equation}
\gamma' = \gamma J_0(\gamma B_{RF}/\omega_{RF})=\gamma J_0(\gamma x),
\end{equation}
where $\gamma$ is the unperturbed gyromagnetic ratio, $B_{RF}$ and
$\omega_{RF}$ are the amplitude and frequency of the applied
oscillating magnetic RF field, $x=\gamma B_{RF}/\omega_{RF}$, and
$J_0$ is the zero-th order Bessel function. Thus by applying an RF
field perpendicular to the static $B_0$ field that satisfies
\begin{equation}
\gamma_n J_0(\gamma_n x) = \gamma_3 J_0(\gamma_3 x),
\end{equation}
the neutrons and $^3$He atoms can be made to precess at the same
frequency (the subscripts $n$ and $3$ refer to the neutrons and the
$^3$He atoms respectively). This condition is called ``critical
dressing.'' In practice, one adjusts the dressing RF field parameter
$x$ by eliminating the oscillating component of the scintillation
light. The signature of a neutron EDM would appear as the change in
$x$ for critical dressing corresponding to the reversal of the
$\bm{E}$ relative to $\bm{B}$. An EDM experiment insensitive to
background magnetic field can thus be performed.

\subsection{Measurement Cycle}
\begin{figure}
\begin{center}
\includegraphics[bb=100 50 500 750,angle=90,width=11cm]{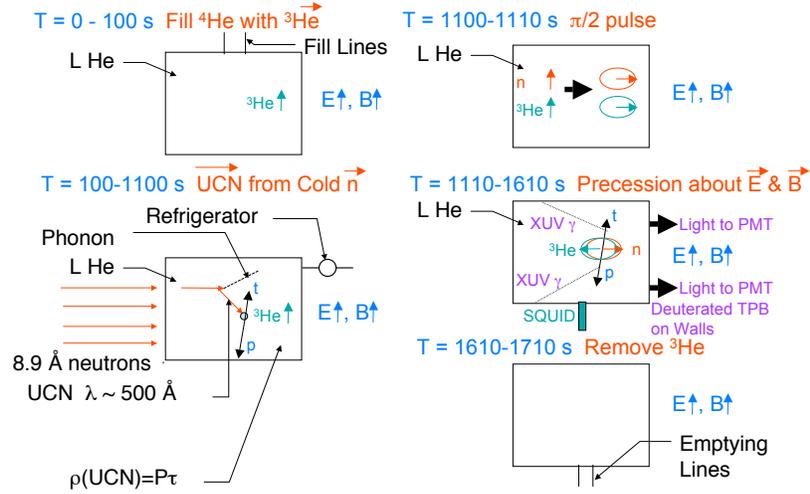}
\end{center}
\caption{Illustrative description of a proposed measurement
  cycle. The duration of each step remains to be optimized to achieve
  maximal sensitivity once the UCN storage and $^3$He relaxation times
  in the measurement cell is known.}
\label{fig:MeasurementCycle}
\end{figure}

To summarize the method described above, a proposed measurement cycle
is illustrated in Fig.~\ref{fig:MeasurementCycle}. Note that the
duration of each step remains to be optimized to achieve maximal
sensitivity once the UCN storage and $^3$He relaxation times in the
measurement cell are known.

\section{FNPB at SNS} 
\label{sec:fnpb}
The experiment will be performed at the Fundamental Neutron Physics
Beamline (FNPB) at the Spallation Neutron Source (SNS) at the Oak
Ridge National Laboratory.  The Spallation Neutron Source (SNS),
currently under construction at the Oak Ridge National Laboratory, is
an accelerator-based neutron source, and will provide the world's most
intense pulsed neutron beams for scientific research and industrial
development~\cite{SNS}.  The Fundamental Neutron Physics Beamline
(FNPB), one of the 24 neutron beamlines in the SNS target hall, is
dedicated to fundamental physics using cold and ultracold
neutrons. 

The FNPB has two neutron beamlines, the ``Cold Neutron Line'' and 
the 8.9~\AA\ Line (or ``UCN Line''). The 8.9~\AA\ Line is dedicated
to experiments that will uses the superthermal process in superfluid
liquid helium to produce ultra-cold neutrons. The 8.9~\AA\ neutrons
will be selected by a double crystal monochromator and will be sent to
an external building located about 30~m downstream, where the nEDM
Experiment will be mounted. 

The construction of SNS completed in 2006. Currently the beam power is
being ramped up towards its full-power capacity of 1.4~MW. The FNPB is
currently under construction. The construction is planned to be completed
in early 2010.

\section{Experimental Apparatus}
\begin{figure}
\begin{center}
\includegraphics[width=12cm]{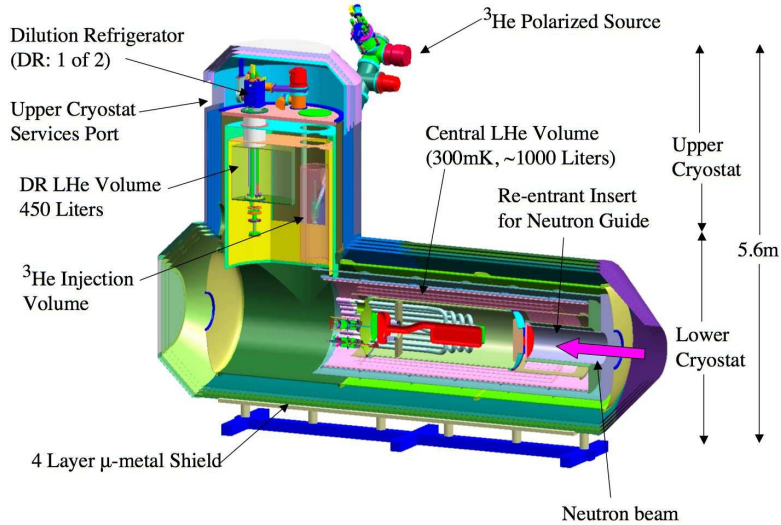}
\end{center}
\caption{The schematic overview of the full detector
apparatus.\label{fig:Apparatus}}
\end{figure}

The conceptual design of the proposed apparatus is shown in
Fig.~\ref{fig:Apparatus}. The picture is derived from a full 3-D
engineering model that has been created to study whether all the
scientific ideas can be realized in single piece of equipment. The
apparatus is divided into two parts, the lower cryostat where the
measurement is made and the upper cryostat where the $^3$He is
injected and removed as well as where the refrigeration is done. Both
lower and upper cryostats are surrounded by four layers of $\mu$-metal
magnetic shields to shield the apparatus from the ambient magnetic
field and its temporal change. Two neutron guides (not shown), which
provide polarized 8.9~\AA\ neutrons, enter from the right and
terminate roughly 50~cm upstream of the high voltage plates.

\begin{figure}
\begin{center}
\includegraphics[width=12cm]{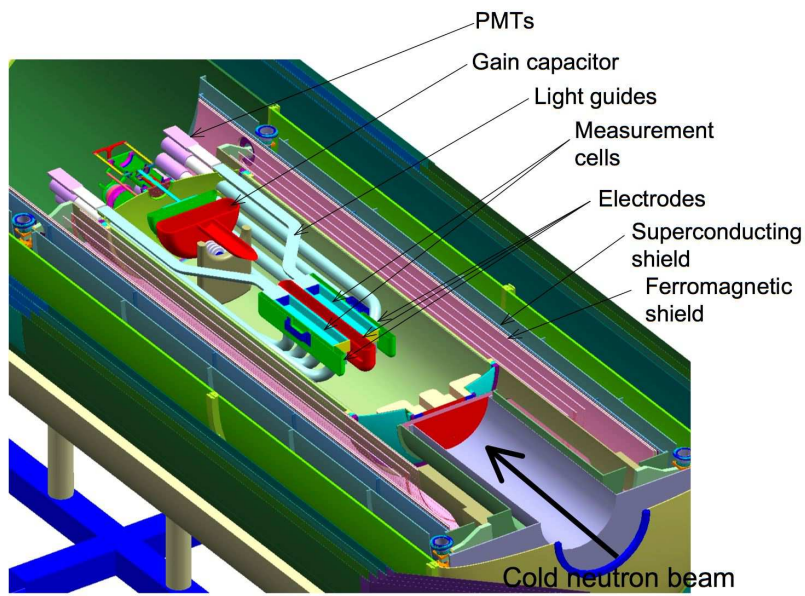}
\end{center}
\caption{The central region of the detector illustrating the
  measurement cells, electrodes, light guides, HV generator and
  magnetic shields.\label{fig:Apparatus2}}
\end{figure}

The cutaway view of the lower cryostat is shown in
Fig.~\ref{fig:Apparatus2}. There are two measurement cells, which are
placed in the gaps between the high voltage and ground electrodes. The
cells are made of acrylic, and the inner walls are coated with
deuterated polystyrene to minimize neutron absorption by hydrogen. The
deuterated polystyrene is loaded with deuterated tetraphenyl
butadinene (dTPB) which serve as the wavelength shifter for the hard
ultraviolet scintillation light from the neutron absorption on
$^3$He. The converted light is guided through the light guides and is
detected by PMTs.  

The high voltage of about 350~kV is necessary to generate the 50~kV/cm
electric field across the 7~cm wide measurement cells. 
A novel technique to supply HV without need for direct 350~kV
application is being developed but we are also developing a high
voltage feedthrough for direct application of the necessary voltage.

The measurement cells, the light guides, the electrodes, and the
variable capacitor are all immersed in a 1200-liter bath of superfluid
helium which serve as the insulator for the high voltage system. Also
the superfluid helium bath surrounding the cells help keep the
temperature across the cells uniform, and eliminate potential heat
sources. This is important because a heat source can generate a phonon
wind which blows away $^3$He atoms thus generating a non-uniformity in
$^3$He concentration. The SQUID magnetometers are mounted on the
ground plates.

A superconducting magnetic shield and a ferromagnetic shield provide a
magnetic shielding in addition to the conventional ($\mu$-metal) room
temperature shields. Located between the ferromagnetic shield and the
1200-liter insulating helium volume are various coils that generate a
uniform static 10~mG field ($B_0$ field), RF pulses to rotate the
spins of the neutrons and $^3$He ($\pi/2$ pulse), and RF pulses for
the dressing field.

\section{Status and Plans}
Currently, the collaboration is refining the conceptual design as well
as vigorously pursuing various R\&D
studies [$15-25$]
that are necessary to optimize the final design of the experiment. The
ongoing and complete R\&D topics include (but are not limited to):
\begin{itemize}
\item study of the neutron storage time in an acrylic cell coated with
  deuterated polystyrene
\item study of dielectric properties of superfluid
  helium~\cite{LON06}
\item theoretical and experimental study of the possible systematic
  effects due to the interference between the motional magnetic field
  effects and the gradient of the $B_0$ field~\cite{LAM05,BAR06}
\item modeling and prototyping of various coils
\item study of the neutron beam line~\cite{ITO06}
\end{itemize} 

The R\&D and the design will continue through calendar year 2007. The
construction is expected to start subsequently. The operation of the
apparatus is expected to start around 2013.

\end{document}